\documentclass{iau} 
\usepackage{graphicx}

\title[Studying halos with future facilities] 
{Studying stellar halos with future facilities}

\author[L. Greggio, R. Falomo  \& M. Uslenghi]   
{Laura Greggio$^1$, Renato Falomo$^1$
 \and Michela Uslenghi$^2$}
 \affiliation{$^1$INAF, Osservatorio Astronomico di Padova,\\Vicolo dell'Osservatorio 5, 35122 Padova, Italy \\ emai:{\tt laura.greggio@oapd.inaf.it, renato.falomo@oapd.inaf.it}\\[\affilskip]
$^2$INAF, Istituto di Astrofisica Spaziale e Fisica Cosmica,\\Via Bassini 15, 20133 Milano, Italy}

\pubyear{2015}
\volume{317}  
\jname{The General Assembly of Galaxy Halos: \\ Structure, Origin and Evolution}
\editors{A. Bragaglia, M. Arnaboldi, M. Rejkuba \& D. Romano, eds.}
\begin{document}

\maketitle

\begin{abstract}
Stellar halos around galaxies retain fundamental evidence of  the processes which lead to their build up.  Sophisticated models of galaxy formation in a cosmological context yield quantitative predictions about  various observable characteristics, including the amount of substructure, the slope of radial mass profiles and three dimensional shapes, and the properties of the stellar populations in the halos. The comparison of such models with the observations provides constraints on the general picture of galaxy formation in the hierarchical Universe, as well as on the physical processes taking place in the halos formation.  With the current observing facilities, stellar halos can be effectively probed  only for a limited number of  nearby galaxies. In this paper we illustrate  the progress that we expect in this field  with the future ground based large aperture telescopes  (E-ELT) and with space based facilities as JWST. 
\end{abstract}

\firstsection 
\section{Introduction}


Nowadays there is plenty of evidence for the presence of streams and substructures in the halo of galaxies, as is expected in the hierarchical models of galaxy formation  (e.g. \cite[Atkinson, Abraham, \& Ferguson 2013]{atkinson13}).
These features, detected down to very low surface brightness ($\mu_V \sim$ 29 mag/arcsec$^2$), are hard to measure against the sky background. Even more difficult is characterizing their stellar populations, estimating total magnitude and color, which trace respectively the mass and the age/metallicity of their stars. In recent years galaxy formation models in a cosmological context have become very sophisticated, yielding detailed predictions about the structure of stellar halos, the amount of substructure, the shapes of shells and streams, the characteristics of the stellar populations in the different components, and more (see \cite[Johnston et al. 2008] {johnston08}; \cite[Font et al. 2011]{font11}; \cite[Cooper et al. 2013]{cooper13}; \cite[Pillepich et al. 2014]{pillepich14}).  The predictions depend on the techniques adopted to construct the models (e.g. whether full hydrodynamical simulations, or n-body models with particle tagging), as well as on parameters describing the physical processes which occur in the troubled galaxy life (e.g. star formation (SF), the initial mass function, the feedback).   
Although the ubiquitous presence of substructures in galaxy halos  qualitatively supports these models, we need to perform a quantitative comparison between predictions and observations, especially  to constrain the models'  parameters. This can be done in different ways, among which
(i) analyzing the demographics of substructures of different types and the properties of their stellar populations; 
(ii) comparing the stellar density profile, checking for the presence and extent of an \textit{in situ} component; 
(iii) measuring the metallicity gradient of stellar halos. For example, halos completely built with stars shed by accreted companions will hardly show a metallicity gradient, while according to full hydrodynamical simulations, a sizable metallicity difference between the outer and the inner halo should exist, since the latter hosts heated disk  stars. 
These kind of studies have been done only for nearby galaxies (\cite[Deason, Belorukov \& Evans 2011]{deason11}; \cite[Greggio et al. 2014] {greggio14}; \cite[Ibata et al. 2014]{ibata14}; Crnojevic,
 this volume); with future large aperture telescopes we expect to pursue these issues on more distant objects,  enriching the samples with a greater number and more types of galaxies.  In this paper we illustrate some examples of these future possibilities, concentrating on photometry of  individual sources, stars and Globular Clusters (GC), which trace the halo formation history.
 
\section{Instrumental set up and stellar probes}

Our results are based on simulations of images computed with the AETC \footnote{http://aetc.oapd.inaf.it}  (v3.0) tool, adopting two instrumental set ups:  the NIRcam camera on board of JWST, and the  MICADO \footnote{http://www.mpe.mpg.de/ir/micado} camera as  baseline for the ELTCAM for the E-ELT telescope. The latter configuration exemplifies the more general case of a $\sim$ 30m-class ground based telescope,  working close to the diffraction limit thanks to an efficient adaptive optics module. 

\begin{figure}
\begin{center}
\includegraphics[width=6in]{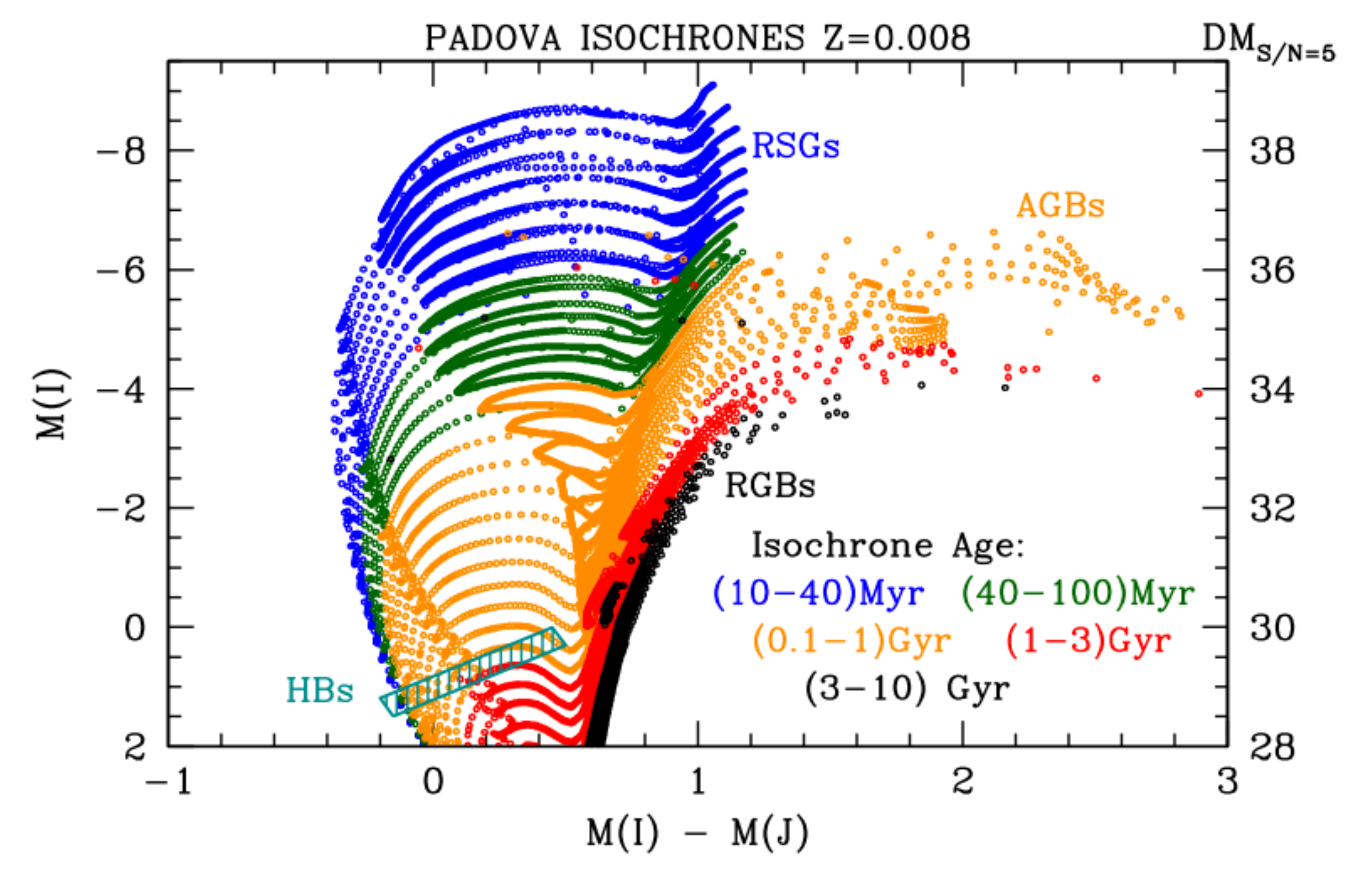} 
\caption{Isochrones computed with the CMD tool (stev.oapd.inaf.it/cgi-bin/cmd) for the indicated ages and metallicity.The right axis is labelled with the distance modulus at which sources of magnitude given on the left axis are measured with S/N=5 in a 3 hrs exposure wit MICADO. The locations of Red Supegiants (RSGs), Asymptotic Giant Branch (AGBs), Horizontal Branch (HBs) and Red Giant Branch (RGBs) stars are indicated.}
\label{fig1}
\end{center}
\end{figure}

The most important features of the MICADO and the NIRcam set ups include the field of view (FoV), respectively of $\simeq 0.78$ and $\simeq 9.33$ arcmin$^2$, the pixel size, of $\sim 3$ and $\sim 32$ mas , and the magnitude limits for an isolated point source, which, in 3 hrs of integration
and at S/N=5, are respectively $I = 30.3$ and 28.8, $J=29.5$ and 28.3. Thus, the NIRcam FoV  is $\gtrsim 10$ times wider than the MICADO FoV, but the latter yields a factor of $\sim$ 10 better resolution. This is important for the photometry in crowded fields, but becomes of little relevance in the low density, halo environment. 
On the other hand, the much larger collecting area of E-ELT allows deeper photometry with MICADO compared to NIRcam, by more than 1 mag.
With these constraints  how far can we detect the stellar probes of the halo formation history?
Fig.\ref{fig1} shows theoretical isochrones  for a wide range of ages. The right  axis, labelled with M(I) + 30.3, indicates the distance modulus up to which the various stellar probes can be measured with good accuracy on a MICADO image.
It appears that bright RSGs will be detected up to very large distances (hundreds of Mpc), but
they trace only the most recent SF. Using AGB stars we will probe the SF of the last few Gyr,  but to  access  the whole SF history we need to sample the bright RGB stars, which can be measured in the galaxy halos up $\sim$ 40 Mpc away. 
The RGB colors are almost insensitive to age, so these stars yield information on the integrated mass of the stellar population, but not on the detailed SF history. On the other hand, they can can be used to evaluate a photometric  metallicity. Finally,  red HB stars can be measured up to $\sim$ 10 Mpc, but blue HB stars can be detected well only within $\lesssim 5$ Mpc. 

In summary, we will be able to catch currently forming stars up to distances of hundreds of Mpcs, which will be very interesting for interacting galaxies. However, in order to
derive the fundamental properties of galaxy halos we need to probe ages of 
several  Gyrs, and the brightest objects we can use to achieve this goal are RGB stars. 

\section{Resolved stars in streams and smooth component in galaxy halos}

\begin{figure}[]
\begin{center}
\includegraphics[width=6in]{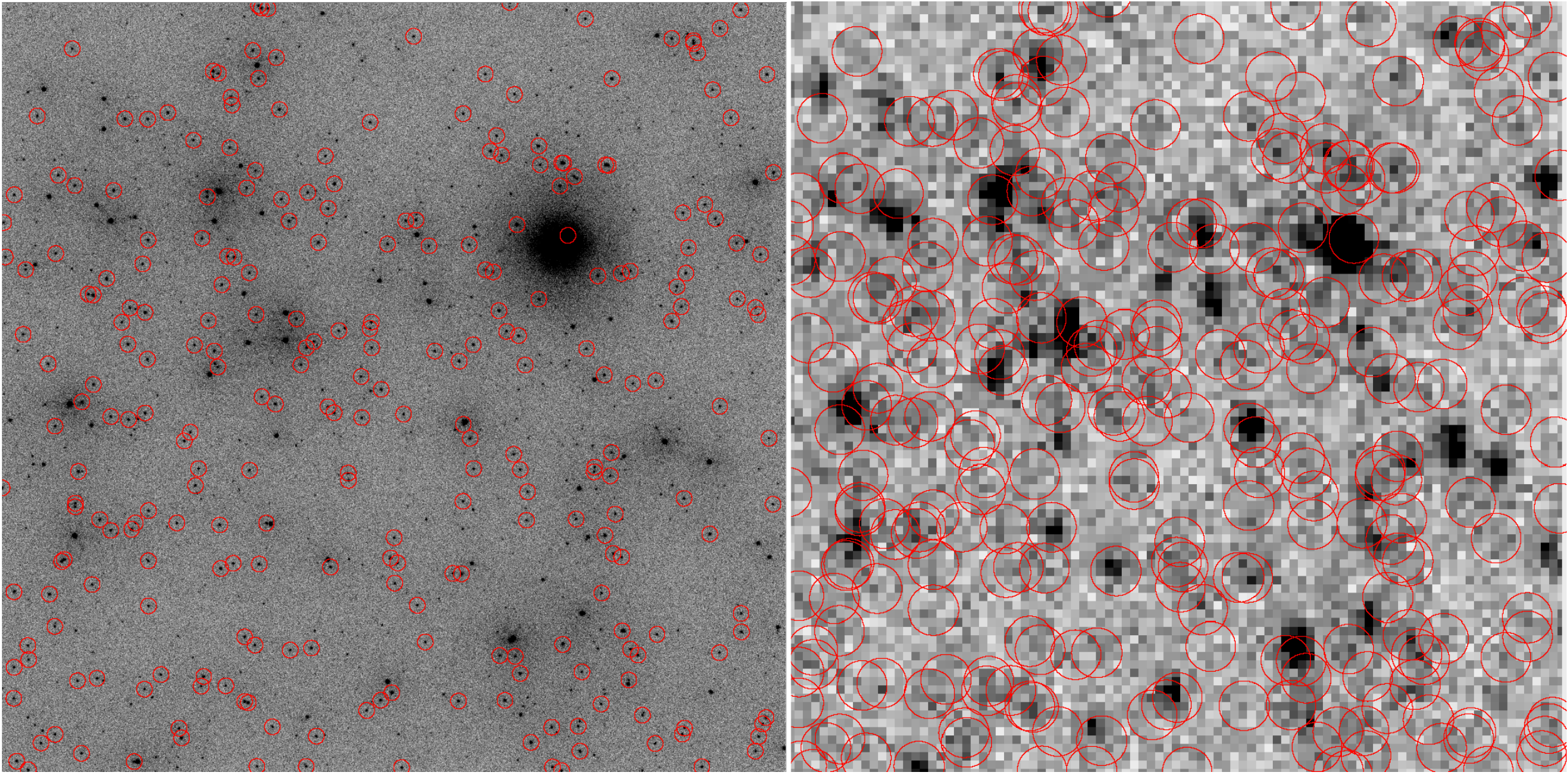} 
\caption{MICADO (left) and NIRcam (right) $3\times3$ arcsec$^2$ $I$-band  simulated images of a stellar population typical for a stream in a galaxy halo at 8 Mpc distance (see text). A total integration time of 3 hrs has been adopted. Circles, with a diameter of 10 (left) and 6 (right) pixels, show the position of red clump stars ($28.5 \leq I \leq 29.5$).}
\label{fig2}
\end{center}
\end{figure}

\begin{figure}[]
\begin{center}
\includegraphics[width=3.5in]{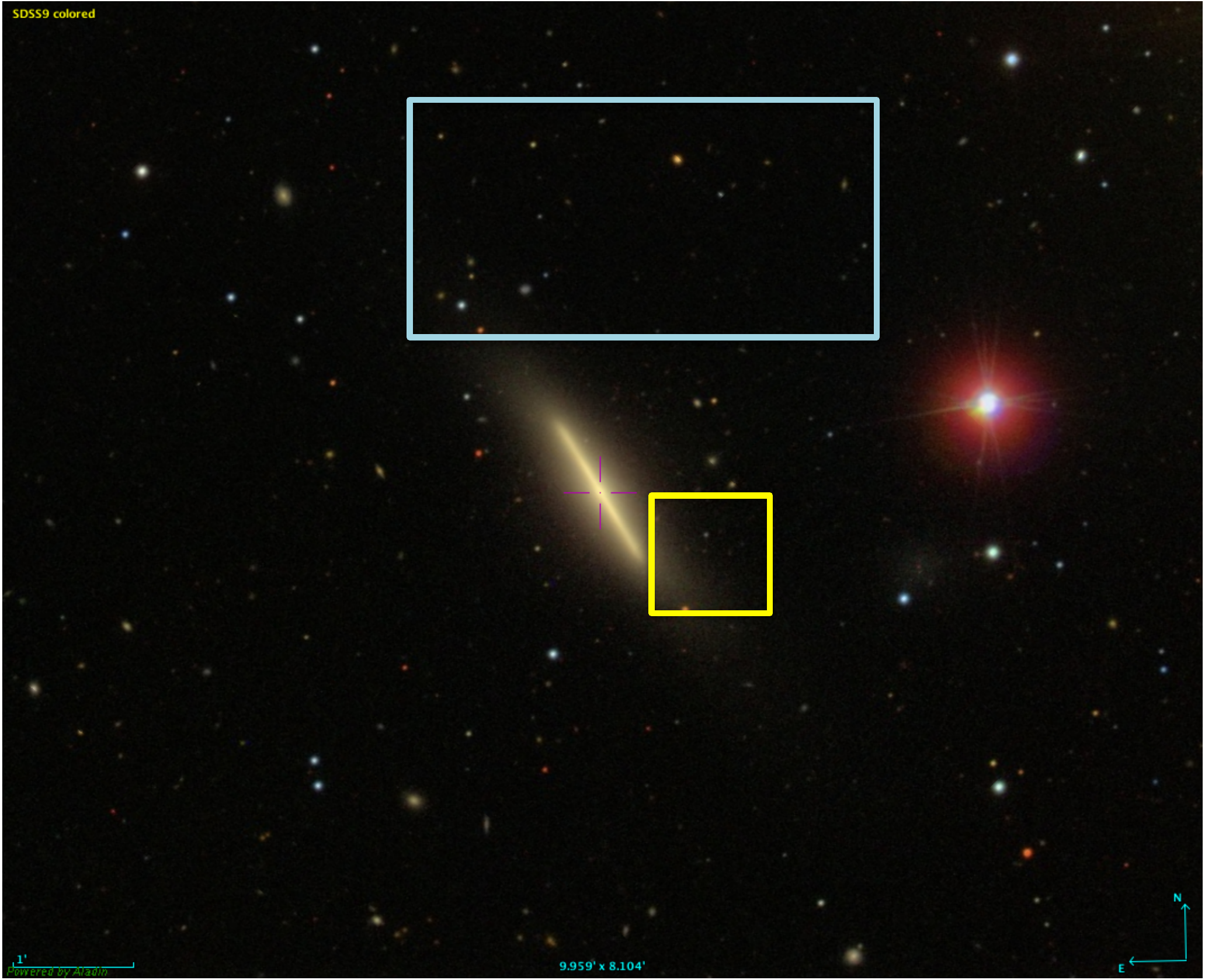} 
 \caption{The lenticular galaxy NGC4452 member of the Virgo Cluster, with superimposed the FoV of NIRcam (cyan) and MICADO (yellow). }
   \label{fig3}
\end{center}
\end{figure}

The characterization of the SF history in the halo substructures and streams yields fundamental information on their progenitors. To illustrate this case we produced simulated images of a synthetic stellar population obtained with a constant SF rate over 10 Gyr, and located at a distance of 8 Mpc, with an average surface brightness of  $\mu_V=24.5$ mag/arcsec$^2$ (Fig. \ref{fig2}).
The luminosity sampled in the field is only of $8 \times 10^4$ L$_{\rm V,\odot}$, and there are 450 stars brighter than $I=30$, but the total MICADO FoV is more than 300 times wider, providing large statistics.  The impressive resolution of the MICADO image is readily appreciated, with faint stars clearly visible through the halo of the bright objects. 
The core helium burning stars (marked with red circles) are too faint and crowded to be measured well on the NIRcam image, while they are very well detected on the MICADO's.  On the other hand, NIRcam will sample well the bright RGB stars, allowing us to measure the density profile,  the shape, and the metallicity of the stellar halo. To better explore this matter we have simulated a 3 hrs integration full NIRcam $I$-band image containing one stellar population with an age of 6 Gyr, metallicity [Fe/H] = -1, and total luminosity of 10$^7$ L$_{\rm V,\odot}$, located at a distance of 8 Mpc, for an average surface brightnes of  $\mu_{V} \simeq$ 28, typical of a galaxy halo. Photometry of the $\simeq 3500$ stars brighter than I = 27.5 (M$_{\rm I} \leq$ -2) shows that they are measured with a mean error of $\lesssim$ 0.04 mag. 
\cite{font11} models for Milky Way type galaxies feature a metallicity gradient with  [Fe/H] $\sim$ -1.2 and -0.5 respectively in  the outer and inner halo regions, which implies  a colour difference of 
$\Delta (I-J) \gtrsim 0.15$ mag between the inner and outer halo bright RGB stars. Comparing this color difference to the photometric accuracy mentioned above it appears that it will be possible to test these models, and in general derive fundamental information on the metallicity of stellar halos of galaxies at this distance. 

The VIRGO cluster of galaxies is the closest rich aggregate of all morphological types.
It is then interesting to examine to which extent we will be able to study the halos of its members. Fig. \ref{fig3} shows a SDSS image of a VIRGO  member galaxy, with superimposed the MICADO and the NIRcam FoV. With 3 hrs integration, the two set ups will allow  sampling the upper 3 and 1.5 mag on the RGB: both limits are quite adequate to derive the photometric metallicity, though MICADO will yield more accurate results.
On the other hand, due to its wide FoV,  NIRcam will be more efficient to map the widespread stellar halo, while MICADO will allow  exquisite photometry in the inner halo regions, at the interface with the disk, at high surface brightness.  

Finally, we can ask how far we can detect substructures in galaxy halos using star counts, arguably the most robust way to trace overdensities at very low surface brightness. Due to the combination of RGB stars color and limiting magnitude, the $J$ band is more efficient than the $I$ band to this end. At half solar metallicity, the tip of the RGB is located at M(J) $\simeq$ -5.5; at the magnitude limits mentioned above for MICADO, we will sample the upper magnitude on the RGB up to  $m-M \sim$ 34.
At such distance, MICADO maps a field of $\sim$ 15 kpc on a side, so we will still need some mosaicking to sample the $\gtrsim$ 100 kpc extension of the halo, but we will be able to detect structures at very low surface brightness.   A stellar population  of 10$^4$ L$_{\rm V,\odot}$  has $\sim 1.4 $ stars in this magnitude range, if its age is older than a few Gyr. 
Thus the surface brightness scales with the surface density $\sigma$ of such stars according to 
$\mu_{\rm V}=-4.8 - 2.5 {\rm log}\, \sigma + m-M$. At $(m-M)=34$,  $\mu_{\rm V} = 33$ corresponds to $\sigma$=0.03 stars/arcsec$^{2}$ which is almost 100 times  higher than the expected density of foregorund stars in this magnitude range, estimated from  the TRILEGAL \footnote{stev.oapd.inaf.it/cgi-bin/trilegal} simulator. Therefore, we expect to be able 
to trace structures to very faint  surface brightness; this is particularly important, since galaxy formation models predict that most substructures are found at $\mu_{V} \gtrsim 30$ (\cite[Johnston et al. 2008]{johnston08}). 

\begin{figure}[]
\begin{center}
\includegraphics[width=6in]{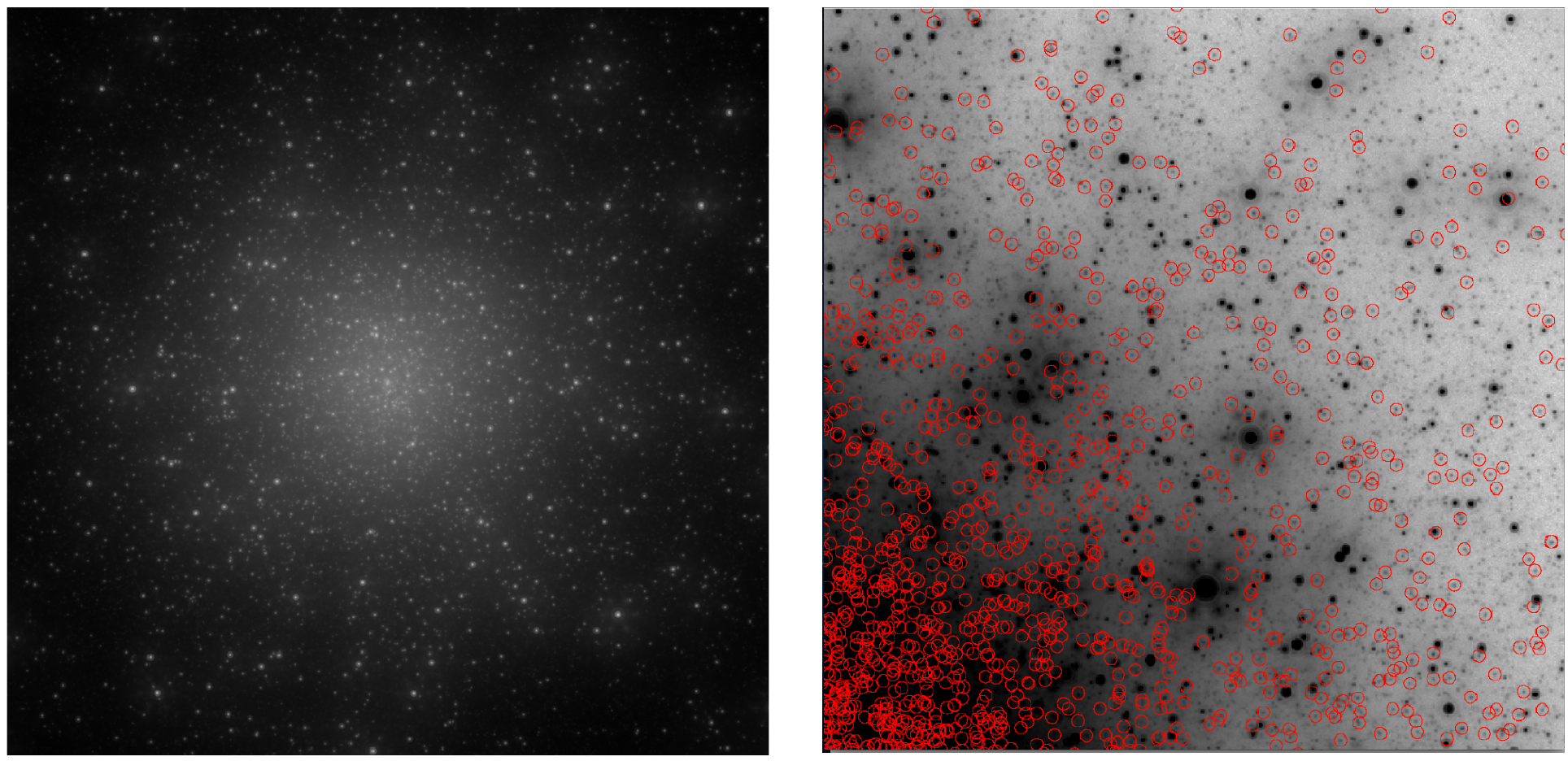} 
 \caption{MICADO simulated image in the $J$-band of a 12 Gyr old stellar population, with [Fe/H]=-1.3, and M(V)=-9.7, adopting a  3 hrs exposure and a distance of 1 Mpc. Stars are distributed
 following a King profile with core and tidal radii of 1 and 30 pc, respectively. On the right panel we show a section of the
 image, extending  from 1.5 pc  to 8.7 pc from the cluster center (lower left and upper right corners), with turn-off stars ($J \simeq 28$) circled in red.}
\label{fig4}
\end{center}
\end{figure}

\section {Globular Clusters and Globular Cluster systems in Galaxies}

One important application for a large aperture telescope working at  the diffraction limit deals with photometry of resolved stars in distant GCs for, e.g.,  their age dating, the detection of multiple populations and  their characterization. To
assess the capabilities in this field a large grid of simulations should be computed and analyzed with sophisticated  photometric tools which allow a full exploitation of the data. Here we just remark that it will be possible to construct Color-Magnitude diagrams down to the turn offs of old clusters in the whole Local Group. Fig. \ref{fig4} shows an example of a simulated image for a GC at 1 Mpc from us.
The stars visible on the left panel are mostly bright giants, but the circled sources on the right panel are turn off stars, with 
$27.8 \leq J \leq 28.2$. These stars, very crowded inside $\sim$ 2 core radii, become well separated in the outer region.
 
Viewed as point sources, GCs hold fundamental information on the formation history of the parent galaxy, as implied by, e.g.,  the relation between the total halo mass and total mass of the GC system (\cite[Hudson, Harris \& Harris 2014]{hudson14}), the bimodal color distribution, the spatial coincidence of clusters and streams  (e.g. \cite[Blom et al. 2014]{blom14}).  Assuming as typical the magnitude distribution of GCs in M31, which peaks at M($J$) = -9.2 (\cite[Nantais et al. 2006]{nantais06}), and using the above mentioned limiting magnitudes for MICADO, we will derive accurate photometry of the GC system in galaxies up to a distance of  $\sim 450$ Mpc, and, thanks to the exquisite resolution, we will detect them even on top of the high surface brightness, central regions. In addition, the large volume accessible with future instrumentation will allow us to characterize the GC systems in hosts of Active Galactic Nuclei (AGNs) of various types and luminosity, and to study the possible connection between the GCs in the host galaxies and the formation of AGNs in their centers, using a sample of hundreds of objects. 
 
\section{Summary and Conclusions}

Based on the results of simulations, we conclude that future telescopes will allow ample and detailed studies of stellar halos  with sufficent accuracy to constrain models of galaxy formation. In particular it will be possible to:
\begin {itemize}
\item catch young stars up to distances of hundreds of Mpcs and witness current stars formation in large samples of interacting galaxies;
\item  costruct the color-magnitude diagram of stars in tidal features down to the red clump for galaxies up to $\sim$ 10 Mpc away, and derive the metallicity distribution and its gradient in the halo of galaxies within 20 Mpc;
\item map the stellar density in galaxy halos up to a distance of 50 Mpc; overdensities will be traceable at very faint limits of surface brightness ($\mu_{\rm V} \sim$ 33
mag/arcsec$^2$);
\item  analyze the color-magnitude diagram of resolved stars in globular clusters down to the old turn-off region  in the whole Local Group;
\item measure the luminosity, mass and color distribution of the globular cluster members of galaxies up to a redshift $z \simeq 0.1$.

\end{itemize}

\begin{acknowledgements}
We acknowledge support from INAF and MIUR through the \textit{Progetto premiale T-REX}.
\end{acknowledgements}

\end{document}